\documentstyle[twocolumn,aps]{revtex}
\input epsf
\epsfverbosetrue
\begin{document}
\draft
\flushbottom
\twocolumn[
\hsize\textwidth\columnwidth\hsize\csname @twocolumnfalse\endcsname

\title{Electron self-trapping in
intermediate-valent SmB$_6$}

\author{S. Curnoe$^1$ and K. A. Kikoin$^2$\footnote{kikoin@bgumail.bgu.ac.il}}

\address{$^1$Weizmann Institute of Science, Rehovot
76100,
Israel\\
$^2$Ben-Gurion University of the Negev, Beer-Sheva 84105, Israel}
\maketitle

\begin{abstract}
SmB$_6$ exhibits intermediate valence in the ground state and
unusual behaviour at low temperatures. The
resistivity and the Hall effect cannot be explained either by
conventional
$sf$-hybridization or by hopping transport in an impurity band.
At least three different energy scales determine three temperature
regimes of electron transport in this system.
We consider the ground state properties, the soft valence fluctuations
and the spectrum of band carriers in n-doped SmB$_6$. The
behaviour of excess conduction electrons in the presence of soft valence
fluctuations and the origin of the three energy scales in the spectrum of
elementary excitations is discussed.
The carriers which determine the
low-temperature transport in this system are self-trapped
electron-polaron complexes rather than simply electrons in an impurity band.
The mechanism of electron trapping is the interaction with soft valence
fluctuations.
\end{abstract}

\pacs{71.28.+d,  71.30.+h, 71.35.Gg, 75.30.Mb}

]
\narrowtext
\tightenlines

\section{Introduction}

Samarium hexaboride is the first compound in which the phenomenon of
intermediate valence (IV) has been seen directly by X-ray absorption
\cite{Bloch65},
but the theory which explains all properties of this compound within a
unified physical model remains elusive in spite of unceasing efforts
of theoreticians and experimentalists during the recent thirty-five years.

The conventional description of the electronic structure of SmB$_{6}$ in
terms
of a two-band promotion model results in a semiconductor-like spectrum with
a
gap $\Delta$. This gap appears as a result of on-site hybridization between
the
narrow band formed by electrons from the samarium $4f$-shell and the wide
conduction band formed by boron $p$-states and samarium $5d$-states.
The theory explains, at least phenomenologically, the main body of
available experimental data \cite{Wacht94}. However, the most
intriguing properties of this material (as well as other IV semiconductors
from the same family, i.e. SmS, TmSe, YbB$_{12}$) cannot be described
adequately within a framework of such a
crude phenomenological picture. Among these
properties are the very nature of the IV ground state, the origin of slow
valence
fluctuations, and the low-temperature anomalies of various physical
properties.
Recent experimental studies of transport
\cite{Bat93,Cool95,Sluchan99}
and optical \cite{Ohta91,Nanba93,Gorsh99} properties revealed the existence
of
several energy scales in the low-energy excitation spectra and several
temperature regimes in the low-temperature electron kinetics.

It is found that at least three different activation energies
influence the behaviour of electrons. 
The biggest energy gap $\Delta_{opt} \approx 14-20$ meV is observed in the
frequency dependent conductivity and dielectric permittivity. 
Apparently,
just this value should be ascribed to the hybridization gap $\Delta$ of the
two-band model mentioned above.
A substantially smaller value of
$\Delta_{act} \approx 3-5$ meV is seen in the low-temperature resistivity
and
Hall effect measurements which display an activation type temperature
dependence in the interval $15  > T > 6$ K. This gap was also measured in
tunnelling experiments \cite{Gunt82}. 
Below 6K the resistivity is nearly temperature independent, and such behavior 
seems to be consistent with hopping transport in an impurity band
separated by the gap $\Delta_{act}$ from the bottom of conduction band. 
However, there is 
still no experimental consensus about the energy scales which control hopping.  
In ref. \cite{Gorsh99} part of the low-T interval is presumed to be described by
Mott's $T^{1/4}$ law for variable-range hopping with a scaling
energy of $T_0=53$ K \cite{Gorsh99}. In another series of samples an
activated term in the conductivity was observed with an activation temperature
$T_a=2.68$ K for $T < 3 K$ \cite{Bat93}, although the derivation of
an activation energy of the
order of several K from measurements made in the same temperature 
interval looks not too trustful. Besides, the pressure dependence 
of the 
residual resistivity \cite{Cool95} is extremely strong, and it hardly can be 
fitted in a picture of non-interacting electrons in an impurity band.
In any case, there is no room for additional excitation branches 
in the mean-field two-band theory, and adding extra localized states in 
the gap does not improve the situation. The standard hybridization model
seems to be too simplistic, and we believe that the generic properties of the
intermediate
valence state can be described only within a framework that goes beyond the
mean field 
approximation.  

In the present paper
we offer a description of the low-energy spectrum of IV SmB$_6$ by
treating the phenomenon of intermediate valence in rare-earth
semiconductors in terms of an excitonic dielectric state. This
approach was offered nearly two decades ago \cite{Kik83}, and its
effectiveness was demonstrated later in explanation of anomalies
in the vibration spectra of IV semiconductors \cite{Kma89}.  Features of
excitonic instability were seen
also in studies of the dielectric-metal transition under pressure
in the compounds TmSe$_{1-x}$S$_x$ \cite{Neun90}.
It is interesting that recently the
dramatic observations of a ferromagnetic phase in $n$-doped divalent
hexaborides CaB$_6$ and SrB$_6$ \cite{Young99} were interpreted in
terms of conventional excitonic instability of a
Keldysh-Kopayev-Volkov-Rusinov 
type \cite{Balva99}.

The theory of intermediate-valence excitonic dielectric
emphasizes the role of soft singlet excitonic states in the
formation of the ground IV state and valence fluctuations in the
excitation spectrum. The IV state is described as a mixture of
singlet $^7F_0$ states of divalent Sm(f$^6$) and the bound
electron-hole pairs $f^5\tilde{b}$ where $\tilde{b}$ is the state
of an electron promoted from the $f$-shell to $p$-orbitals spread
over neighbouring boron sites but having the same symmetry as the
f-electron in a central site. In some sense these states are
electron-hole analogs of Zhang-Rice two-hole states known in the
theory of low-energy states of CuO$_2$ planes in high-T$_c$
perovskites \cite{Zhang88}.

Here we consider the case of doped $n$-SmB$_6$ material. We study the
behaviour of excess conduction electrons in the presence of soft valence
fluctuations and discuss the origin of three energy scales seen in
different temperature regimes. We show that the carriers which determine the
low-temperature transport properties in this system are self-trapped
electron-polaron complexes rather than simply electrons in an impurity
band.
The mechanism of electron trapping is the interaction with soft valence
fluctuations.

\section{Ground state of intermediate valence semiconductor}

SmB$_{6}$ together with ``golden'' SmS is considered to be a classical
example
of a non-magnetic intermediate
valence (IV) semiconductor (see \cite{Wacht94} and
references therein). Both of these compounds possess a singlet ground state
with intermediate valence $v\sim +2.6$ . The main difference in the
properties
of these two compounds is that SmS transforms into an IV semiconductor only
at
finite pressure (or under the chemical pressure of rare earth
ions of smaller radius
substituting for Sm) whereas SmB$_{6}$ possesses intermediate valence and
concomitant anomalous properties at ambient pressure.

Apparently, differences in the electron band spectra of these two materials
are the eventual source of differences in their properties. SmS at ambient
pressure is known to be a normal semiconductor with a relatively wide gap in
the energy spectrum $\Delta =0.23$eV.
This gap divides the filled $f$-levels of  Sm$^{2+}(f^{6})$ ions from the
bottom of conduction band formed mainly by Sm $d$-states.
The $4f$-levels, in turn,  form a nearly dispersionless valence band within
a
wide forbidden gap $E_g \gg \Delta$
between the conduction band and conventional valence band formed
mainly by $p$-electrons of the chalcogen sublattice.
At finite pressure
the gap $\Delta$ is suppressed, but instead of collapsing to zero
it transforms into a microgap $\Delta_\mu$
at some critical pressure $P_c$,  and the
material acquires the properties of an IV semiconductor.
On the other hand, 
the ``parent" divalent hexaborides without a $4f$-shell (CaB$_6$, SrB$_6$),
as
well as trivalent LaB$_6$ with an empty $4f$-shell, have a band
structure
without a gap in the actual energy interval. The density of
states
near the Fermi level in hexaborides 
 is predetermined by a single degenerate
band with a minimum in the $X$-point of the Brillouin zone. This band is
formed mainly by $p$-electrons of the boron sublattice with a small
admixture of
$d$-states of the cation sublattice. It is nearly empty in divalent
semimetallic hexaborides \cite{Has79,Mas97} and more than half-filled in
trivalent LaB$_6$ \cite{Har88}. Thus, one can expect that the $f$-level
which is responsible for the intermediate valence of SmB$_6$ should cross
this simple band. Indeed, conventional band calculations \cite{Yan92}
give a band structure compatible with this presumption.

Despite the differences between the parent (unhybridized) spectra
of SmS and SmB$_6$, the decisive
similarity of these two cases is the closeness or overlapping of the highly
correlated $4f$-levels of Sm$^{2+}$(4f$^6$) and the nondegenerate conduction
band. In SmB$_6$ these states overlap at ambient pressure, whereas
in SmS the closeness is
achieved under external pressure which induces
the excitonic instability \cite{Kik83} at $P_{c}$.
The closeness results in promotion of electrons to delocalized band
states. Crucial is the fact that the number $y$ of promoted electrons is
the deviation from the divalent state in both Sm$^{(2+y)+}$S and
Sm$^{(2+y)+}$B$_6$ IV semiconductors. According to the conventional band
scheme
which treats the $ff$-interaction in a self-consistent mean-field
approximation, the $fp$-mixing results in the appearance of a hybridization
gap
$\Delta$ in case of SmB$_6$ and a pseudogap in the mixed valent phase 
of SmS. To take into account
strong on-site correlations, this scheme should be modified.

The starting Hamiltonian for treating IV semiconductors is the
Anderson lattice Hamiltonian supplemented by an interband Coulomb
interaction
(see, e.g., \cite{Martin79}):
\begin{eqnarray}
H&=&\sum_{{\bf k}\sigma}\epsilon_{\bf k} a_{{\bf k}\sigma
}^{\dagger}a_{{\bf k}\sigma }+ \sum_{\bf
m}\sum_{\Lambda=0,\{\Gamma \mu\}} E_\Lambda |{\bf
m}\Lambda\rangle\langle{\bf m}\Lambda|  \nonumber \\ 
&+&\sum_{{\bf
mk}\sigma\{\Gamma \mu\}}\left[V_{{\bf mk}\sigma,\Gamma \mu}
a_{{\bf k}\sigma }^{\dagger}|{\bf m}\Gamma \mu\rangle\langle{\bf
m}0| + H.c. \right] \nonumber \\
&+&\sum_{\bf jkk'\sigma}U_{\bf mkk'}n_{{\bf
m}f}a_{{\bf k}\sigma }^{\dagger}a_{{\bf k'}\sigma } . \label{1.0}
\end{eqnarray}
Here $\epsilon_{\bf k}$ is a simple spin-degenerate dispersion law
for the band electrons (we assume for the sake of simplicity that
these electrons come from the boron $2p$-shells). The states
$|{\bf m}\Lambda\rangle$ of the Sm ion at site ${\bf m}$ are
represented by two configurations, $|{\bf m}\Lambda\rangle=|{\bf m
}0\rangle \equiv |{\bf m}^7F_0\rangle$ for the divalent state
4f$^6$ and $|{\bf m}\Lambda\rangle=|{\bf m}\Gamma \mu\rangle$ for
the trivalent state 4f$^5$. $\Gamma$ stands for the multiplet
$^6H_{5/2}$ of the 4f$^5$ configuration which splits into a
$\Gamma_7 \mu$ doublet and a $\Gamma_8 \mu$ quartet by the cubic
crystal field $\Delta_{CF}$; $\mu$ enumerates the states of
corresponding irreducible representations, and $n_{{\bf m}f}=
\sum_\mu|{\bf m}\Gamma \mu\rangle \langle{\bf m}\Gamma \mu|$ is
the $f$-electron occupation number. The hybridization interaction
describes the promotion of electrons from the $f$-shell to the
conduction band accompanied by a change of atomic configuration.
The strong interband Coulomb interaction is given by $U_{\bf
mkk^{'}}$.  The hybridization matrix element is $V_{{\bf
mk}\sigma,\Gamma \mu}= \langle {\bf k}\sigma,{\bf m}\Gamma
\mu|H|{\bf m}0\rangle$. At $T\ll \Delta_{CF}$ only the lowest
doublet state $\Gamma_7$ should be taken into account, so one
should deal with hybridization of two doubly degenerate states.

The strong intrashell interaction is inserted in the configuration
change operators $X_{\bf m}^{\Lambda \Lambda'}=|{\bf
m}\Lambda\rangle\langle{\bf m}\Lambda'|$. To make the
hybridization problem solvable the nondiagonal operators are
usually represented in a factorized form $X_{\bf m}^{\Gamma 0} =
f_{{\bf m}\mu} b^\dagger_{\bf m}$ (see, e.g., \cite{Barn76}),
where $f_{{\bf m}\mu}$ and $b^\dagger_{\bf m}$ describe auxiliary
fermion and boson fields which correspond to charge and spin
degrees of freedom. This procedure gives reasonable results in the
case of nearly integer valence, but in the IV state the spin
charge separation procedure can hardly be useful even as a
starting approximation.  We prefer to use another approach
\cite{Kik80} which seems to be adequate in the nonmetallic case
when one can hope that hybridization will result in the separation
of strongly correlated and weakly correlated bands in the energy
space.

According to the prescription of this approach, we pick out the
one-electron band Hamiltonian $H_b$ which describes the hybridization
of band electrons with the mean-field level $\varepsilon_{f}$, which is
defined as the energy difference
$$
\varepsilon_{f}=E_0-E_{\Gamma_7}
$$
where $E_{\Gamma_7}$ is the lowest (doubly degenerate) state
in the multiplet f$^5(^6H_{5/2}$), and then represent the Hamiltonian
(\ref{1.0}) in the form $H=H_b+\Delta H$. The term $\Delta H$
describes the excitations above the ground state.
Then the mean field
spectrum in the actual energy interval can be described approximately by a
Hamiltonian
\begin{equation}
H_{b}=\sum_{s=1,2}\sum_{{\bf k}\sigma}\varepsilon _{s}({\bf k})
c_{s,{\bf k}\sigma }^{\dagger}c_{s,{\bf k}\sigma }~.
\label{1.1}
\end{equation}
Here
\begin{equation}
\varepsilon _{1,2}({\bf k})=
\frac{1}{2}(\epsilon_{\bf k}+\varepsilon _{f})\mp
\sqrt{\frac{1}{4}(\epsilon_{\bf k}-\varepsilon _{f})^{2}+|V({\bf k})|^{2}},
\label{1.2}
\end{equation}
where $e^{i{\bf m}\cdot{\bf k}}V({\bf k}) = V_{\bf mk}$ from Eq.
(\ref{1.0}). Hybridization between $p$- and $f$-states due to the
matrix elements $V({\bf k})$ results in the wavefunctions
\begin{eqnarray}
c_{1,{\bf k}\sigma } =u_{\bf k}f_{\sigma }+
v_{\bf k}c_{{\bf k}\sigma }
\label{1.4} \\
c_{2,{\bf k}\sigma } =-u_{{\bf k}\sigma }c_{{\bf k}\sigma }+
v_{{\bf k}\sigma }f_{\sigma }  \nonumber
\end{eqnarray}
where the coefficients of the canonical transformation are
\begin{eqnarray}
u_{\bf k}&=&\frac{1}{2}\left[
1+\frac{\epsilon_{\bf k}-\varepsilon _{f}}
{\sqrt{(\epsilon_{\bf k}-\varepsilon _{f})^{2}+4|V({\bf k})|^{2}}}
\right]\nonumber \\
v_{\bf k}^2&=&1-u_{\bf k}^2.
\label{1.4a}
\end{eqnarray}

Since the band states
and the $f$-states have the same symmetry at the points $X_{7}$, $R_{7}$
$\Gamma_{7}$ and $\Delta{_7}$ \cite{Martin79} a hybridization gap opens,
and the lower band $\varepsilon_{1}$ is filled in the ground state of
``divalent" SmB$_6$ because two electrons transferred from the
$f_{\Gamma_{7}}$
level to the (initially empty) band $\varepsilon_1$ fill it completely.
Let us assume that the level $\varepsilon_f$ crosses the band
$\epsilon({\bf k})$ in its lower part, so that the band
$\varepsilon_1({\bf k})$ has mainly $f$-character. Then the square root
in Eq. (\ref{1.2}) can be expanded in
$V({\bf k})/[\varepsilon_{\bf k}-\varepsilon _{f}]$ for most of
the Brillouin zone, and the Wannier states given by the operators
$c_{1,{\bf m}\sigma}=
N^{-1/2}\sum_{\bf k}\exp (i{\bf k}\cdot {\bf m})c_{1,{\bf k}\sigma}$
can be approximated by the following equations,
\begin{eqnarray}
&&c_{1,{\bf m}\sigma}\approx f_{{\bf m}\sigma}\nonumber \\
&&\hspace{.2in}+
N^{-1/2}\sum_{\bf k}\sum_{\langle{\bf j}\rangle_{NN}}
\frac{V({\bf m-j})\exp [i{\bf k}\cdot ({\bf m-j})]}
{\epsilon_{\bf k}-\varepsilon _{f}}a_{\bf j}.
\label{1.4b}
\end{eqnarray}
Here $\langle{\bf j}\rangle_{NN}$ are the nearest neighbours (boron sites)
of a Sm ion in a given
crystal cell and $V({\bf m}-{\bf j})$ is the Fourier transform of V({\bf
k)}.
The width $T$ of the lower band is determined by a hopping
integral
which can be estimated as
\begin{equation}
T_{\bf mn}\approx N^{-1/2}\sum_{\bf kj} \frac{V({\bf m}-{\bf j})
V^*({\bf n}-{\bf j})\exp [i{\bf k}\cdot ({\bf m}-{\bf n})]}
{\epsilon_{\bf k}-\varepsilon _{f}}
\label{1.4c}
\end{equation}
(here ${\bf j}$ are the nearest neighbors of both ${\bf m}$ and
${\bf n}$).

 To study  the {\it real} carriers, one should go beyond the
mean-field Hamiltonian, i.e. take into account the term $\Delta H$
which includes, in particular, the interaction between holes in
the narrow band $\varepsilon_1({\bf k})$ and the electron-hole
interaction between carriers in different bands. The former
contribution stems from a strong  Coulomb interaction between
electrons in the Sm $f$-shell $U=\langle f_{\bf m}f_{\bf
m}|U|f_{\bf m}f_{\bf m}\rangle$. This Hubbard interaction is
reduced due to slight delocalization of Wannier functions
$$
\tilde{U}=\langle c_{1,\bf m}c_{1,\bf m}|U|c_{1,\bf m}c_{1,\bf m}\rangle
=r^4 U,
$$
where $r=\int
S(\epsilon)u(\epsilon)d\epsilon$ is the reduction (nephelauxetic)
factor \cite{Kik80}. This factor depends on the density of states
$S(\epsilon)$ in the unhybridized band $\epsilon_{\bf k}$ (see Eq.
\ref{1.0}), and the coefficient $u(\epsilon)$ is that which
appears in Eq. (\ref{1.4}) but written as a function of energy. In
our case the condition $T\ll \tilde{U}$ is assumed to be valid,
and this means that no more than a single hole per site can be
created in this narrow ``Hubbard" band.

The upper band $\varepsilon_2({\bf k})$ is formed mainly by
$p$-electrons with admixture of an $f$-component in the states
close to the bottom of this band. If the bottom of the conduction band
is close to the center of the Brillouin zone \cite{Martin79}, the
Bogolyubov's coefficients for the electron wavefunction described
by the operator $c_{2,\bf k}$ (\ref{1.4}) are given by
\begin{equation}
v_k^2\approx 1-\frac{|V_0|^2}{(\varepsilon _{f}-\epsilon_{\bf
k})^2},\;\;\;u_k\approx \frac{V_0}{\varepsilon _{f}-\epsilon_{\bf
k}}. \label{1.4d}
\end{equation}
The Wannier operators $c_{2,\bf m}$ have the same
form as (\ref{1.4b}), except that the sign of the denominator in the
second term is negative.  The dispersion in this band can be
approximated by the effective mass law
\begin{equation}
\varepsilon_{2}(k)\approx \Delta_0+\frac{\hbar^2(k-k_b)^2}{2m^{*}_c}
\label{1.2a}
\end{equation}
where $\Delta_0$ is
the gap in the two-band spectrum (\ref{1.2}).
Here $k_{b}$ is the wavevector corresponding to the bottom of the
empty band, and the effective mass can be estimated as 
\begin{equation}
m^*_c/m_0  \approx
[(\varepsilon_f-\epsilon_{k_b})/V(k_b)]^2 = 1/\xi_c ~.
\label{1.2b}
\end{equation}
The effective mass
is noticeably heavier than the bare mass $m_0$ of the conduction
electron (experimentally, $m^*_c\sim 100 m_0$ \cite{Gorsh99}). We
suppose that the hybridization gap is seen as the biggest gap
$\Delta_{opt}$ in the experiments mentioned in the Introduction.

Now, taking the energy of the state with filled lower band as a
reference
point in our calculations of the excitation energy spectrum, we have the
effective Hamiltonian
\begin{eqnarray}
\widetilde{H} &=&
\sum_{{\bf mn},\sigma}t_{\bf mn}\tilde{c}^\dagger_{1,{\bf m}\sigma}
\tilde{c}_{1,{\bf n}\sigma}+
\frac{\tilde{U}}{2}\sum_{{\bf m}\sigma}n_{1,{\bf m}\sigma}n_{1,{\bf
m}-\sigma}\nonumber \\
&&+
\sum_{{\bf k}\sigma }\varepsilon _{2}(k)
c_{2,{\bf k}\sigma }^{\dagger}c_{2,{\bf k}\sigma }
+ H_{12}
\label{1.5}
\end{eqnarray}
where $t_{\bf mn} = \sum_{\bf k} e^{i{\bf k}\cdot({\bf m}-{\bf n})}
\varepsilon_1({\bf k})$ and $\varepsilon_{1,2}({\bf k})$ are given by
Eq. (\ref{1.2}).
The electron states in the hopping term for hole excitations
are dressed in projection operators,
$\tilde{c}_{1,{\bf m}\sigma} = c_{1,{\bf m}\sigma}n_{1,{\bf m}-\sigma}$.
This
projection inserts the kinematic restriction for the hole motion:
the hole can be created at a given site only provided the electron
with the opposite spin is still at the same site. Therefore, the holes
are practically immobile.
The last term $H_{12}$ is responsible for the electron-hole interaction,
\begin{equation}
H_{12}=\sum_{{\bf k}_{1}{\bf k}_{2}{\bf k}_{3}{\bf k}_{4}}
\sum_{\sigma \sigma ^{\prime }}W({\bf k}_{1},{\bf k}_{2},{\bf k}_{3},
{\bf k}_{4})c_{1,{\bf k}_{1}\sigma }^{\dagger }c_{1,{\bf k}_{2}
\sigma }c_{2,{\bf k}_{3}\sigma ^{\prime }}^{\dagger }c_{2,{\bf k}_{4}
\sigma ^{\prime }}
\label{1.3}
\end{equation}
where $W$ is derived from the Coulomb interaction [last term in
the Hamiltonian (\ref{1.0})].

The two-band Hamiltonian $\widetilde{H}$ contains hybridisation
built into the electron and hole states, and the average occupation numbers
$\bar{n}_{1{\bf i}f}\equiv \bar{n}_f$ are formally less than one 
in the lower ``Hubbard" band $\varepsilon_1$.
However, it should be emphasized that this deviation from integer value
($\bar{n}_f=1$ corresponds to the Sm$^{2+}$ state) is not
the true IV state. 
The one-electron picture given by
the mean-field Hamiltonian $H_b$ (\ref{1.1}) implies that interband
transitions given by the operator
\begin{equation}
S_q=\sum_{\bf k} c_{2,{\bf k}+{\bf q}\sigma }^{\dagger}c_{1,{\bf k}\sigma}=
N^{-1}\sum_{\bf mj}e^{-i{\bf q}\cdot{\bf m}}
c_{2,{\bf m}+{\bf j}\sigma}^{\dagger}c_{1,{\bf m}\sigma}
\label{1.3a}
\end{equation}
form the fundamental branch of elementary excitations in this
semiconductor. However, free electron-hole excitations alone
cannot explain the unusual properties of IV SmB$_6$. We believe
that in the IV state an extra branch of charge transfer
excitations exists. These are valence fluctuations, and are
responsible for the unusual low-energy electron spectra and the
numerous anomalies of the physical properties of SmB$_6$.
According to the scenario suggested in \cite{Kik83,Stev78} and
verified in \cite{Kma89}, the true IV ground state arises as a
result of admixture of singlet excitonic states to the ground
state of the same symmetry, and this admixture is non-negligible
when the binding energy of the exciton is comparable with the band
gap, i.e. when the system is close to the excitonic instability.
To realize this scenario, one should construct the singlet exciton
for the specific case of SmB$_6$.

\begin{figure}[ht]
\epsfysize=3.2in
\epsfbox[-20 100 469 560]{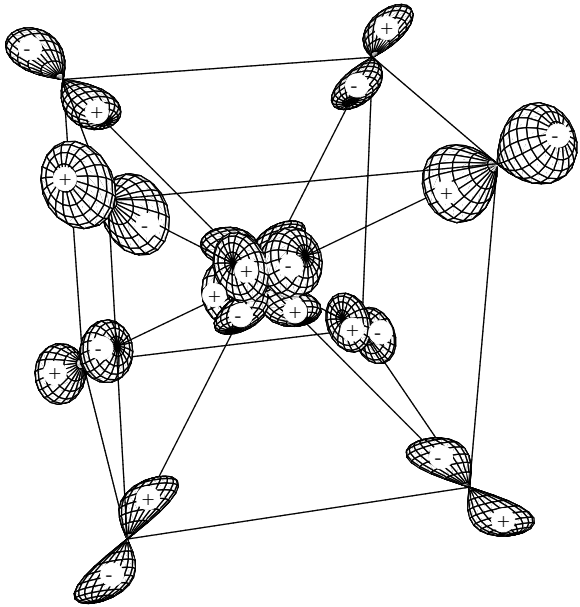}
\caption{A $4f$-orbital located at the centre surrounded by
eight $p$-orbitals on the ($\frac{1}{2}$,$\frac{1}{2}$,$\frac{1}{2}$)
corners.  The signs of the $p$-orbitals are chosen to have the
same symmetry as the $f$-orbital.
\label{fi:fp}}
\end{figure}

According to the theory of IV rare-earth semiconductors \cite{Kik83,Stev78},
the singlet exciton is a bound state of a hole in the samarium
$f$-shell and an electron spread over the $p$-states of the surrounding NN
boron atoms
with the same crystal point symmetry as an $f$-electron in the central
cell . This state is constructed from electron-hole
pairs (\ref{1.3a}) by means of an envelope function $F_{\bf q}({\bf m-j})$.
When ${\bf q}=0$, the exciton operator can be written as
\begin{equation}
|\Psi_{ex}\rangle=
N^{-1/2}\sum_{\bf m}\sum_{\langle{\bf j}\rangle_{NN}}F_{0}({\bf m}-{\bf j})
p_{{\bf m}+{\bf j}\sigma}^{\dagger}f_{{\bf m}\sigma}
|0\rangle ~.
\label{1.7}
\end{equation}
Here a basis of $f$ and $p$-orbitals is chosen. As mentioned above,
the $f$-electron has $\Gamma_7$ symmetry, and its angular dependence is
determined by the $xyz$ cubic harmonics. The envelope amplitude $F_0$ is
a function of exciton energy $E_{ex}$ (see, e.g., Ref. \cite{Kik83} where
this function is calculated in the NN-approximation). It also contains the
phase factor which orders the phases of the $p_x$, $p_y$ and $p_z$ boron
orbitals in accordance with the required symmetry (cf. \cite{Zhang88}),
\begin{equation}
F_{0}({\bf m-j})=F(E_{ex})(-1)^{M_{\bf m,j}}
\label{1.8}
\end{equation}
where the phase $(-1)^{M_{\bf m,j}}$ follows the arrangement shown in Fig.
\ref{fi:fp}.

In fact, it will be
shown below that the IV ground state resembles in some sense the Zhang-Rice
(ZR)
singlets \cite{Zhang88} which are believed to be formed in Cu-O planes of
high-T$_{c}$ materials. The Emery Hamiltonian \cite{Emery88}, which is the
starting point for the description of hybridization between weakly
interacting
oxygen $p$-electrons and strongly correlated copper $d$-holes is similar to
the Anderson Hamiltonian describing the hybridization of nearly free
$b$-electrons
in the conduction band with strongly localized electrons in samarium
$4f$-shells.  The difference is that the ZR bound states of a local Cu spin
and
a hole distributed over $p$-orbitals of the surrounding O ions are formed as
excitations in doped oxicuprates, while the IV singlets are formed in the
ground state as bound state consisting of a
hole in the $f$-shell of the Sm ion and
an electron distributed over the $p$-orbitals of the surrounding B ions of
the
cation sublattice . The binding mechanisms are the antiferromagnetic
$dp$-exchange in ZR case and the Coulomb $fp$-attraction in our case.

Now, inserting (\ref{1.8}) into (\ref{1.7}), we have
\begin{equation}
|\Psi_{ex}\rangle=F(E_{ex})N^{-1/2}\sum_{{\bf m}\sigma}
P^\dagger_{{\bf m}\sigma}f_{{\bf m}\sigma}|0\rangle
\label{1.9}
\end{equation}
where $P^\dagger_{{\bf m}\sigma}$  is a localised state which
is defined as follows.
SmB$_6$ has a simple cubic lattice structure, with an Sm atom in the
centre of each cell, and a B$_6$ group on each corner.  To simplify the
treatment, we approximate the electronic wavefunctions for the boron
clusters
by the $p$-orbital directed along the diagonal of the cube
(see Fig. \ref{fi:fp}). These orbitals are responsible for the
covalent bonding between the boron clusters and Sm sublattice \cite{Has79}.
We introduce the linear combinations
$$p^{xyz}_{\bf j} \equiv \frac{1}{\sqrt{3}}(p_x({\bf j})+p_y({\bf j})+
    p_z({\bf j}))
$$
$$
p^{x\bar{y}z}_{\bf j} \equiv \frac{1}{\sqrt{3}}(p_x({\bf j})-
p_y({\bf j})+p_z({\bf j})),$$
etc. which  represent $p$-type electronic wavefunctions centred
at ${\bf j}$ and oriented in the direction $\hat{x}+\hat{y}+\hat{z}$.
In analogy with the Zhang-Rice construction,  
we write down a
localised state for the Sm site ${\bf m}$ consisting of eight nearest
neighbour $p$-orbitals oriented to have the same symmetry as a central
$f$-orbital, as shown in Fig. \ref{fi:fp},
\begin{eqnarray}
 P_{{\bf m}\sigma}& =&
\frac{1}{\sqrt{8}}\left[
p^{xyz}_{{\bf m}+(\frac{1}{2},\frac{1}{2},\frac{1}{2})}
 +  p^{xyz}_{{\bf m}-(\frac{1}{2},\frac{1}{2},\frac{1}{2})}
 +   p^{x\bar{y}\bar{z}}_{{\bf m}+(-\frac{1}{2},\frac{1}{2},\frac{1}{2})} 
\right. \nonumber \\
&&+ p^{x\bar{y}\bar{z}}_{{\bf m}-(-\frac{1}{2},\frac{1}{2},\frac{1}{2})}
 +   p^{\bar{x}y\bar{z}}_{{\bf m}+(\frac{1}{2},-\frac{1}{2},\frac{1}{2})}
+ p^{\bar{x}y\bar{z}}_{{\bf m}-(\frac{1}{2},-\frac{1}{2},\frac{1}{2})}
\nonumber \\
&&\left.
 +  p^{\bar{x}\bar{y}z}_{{\bf m}+(\frac{1}{2},\frac{1}{2},-\frac{1}{2})} +
p^{\bar{x}\bar{y}z}_{{\bf m}-(\frac{1}{2},\frac{1}{2},-\frac{1}{2})}
\right].
\label{1.10}
\end{eqnarray}

States on neighbouring sites are non-orthogonal because of shared
$p$-electrons.  Therefore, it is useful to make a canonical transformation
to the orthonormalized Wannier or Bloch operators $d^{\dagger}$,
\begin{equation}
P^\dagger_{{\bf n}}=\sum_{\bf m}\lambda({\bf m}-{\bf n})d_{\bf m}^\dagger
=N^{-1/2}\sum_{\bf k}\beta^{1/2}_{\bf k}d^\dagger_{{\bf k}}
e^{-i{\bf k}\cdot{\bf n}},
\label{1.11}
\end{equation}
where
\begin{equation}
\lambda({\bf m}-{\bf n})=
N^{-1}\sum_{\bf k}\beta^{1/2}_{\bf k}e^{i{\bf k}\cdot({\bf m}-{\bf n})},
\label{1.11a}
\end{equation}
\begin{eqnarray}
\beta_{\bf k} & = & \frac{8}{3}\left[\cos^2\frac{k_x}{2}\sin^2\frac{k_y}{2}
        \sin^2\frac{k_z}{2} + \sin^2\frac{k_x}{2}\cos^2\frac{k_y}{2}
        \sin^2\frac{k_z}{2}\right. \nonumber \\
 & &  \left.  +\sin^2\frac{k_x}{2}\sin^2\frac{k_y}{2}
        \cos^2\frac{k_z}{2}\right].
\label{1.12}
\end{eqnarray}
Then the exciton operator acquires the form
\begin{eqnarray}
|\Psi_{ex}\rangle&=&F(E_{ex})\frac{1}{N}\sum_{{\bf m}\sigma}
\left[
\lambda_0 d^\dagger_{{\bf m}\sigma}+\sum_i
\lambda_i
\sum_{\langle {\bf n}\rangle_{i}}
d^\dagger_{{\bf n}\sigma}
\right]f_{{\bf m}\sigma}|0\rangle \nonumber \\
&\equiv&\frac{1}{N}
\sum_{\bf m}|\psi_{\bf m}\rangle\;.
\label{1.13}
\end{eqnarray}
Here the index $i=1,2\ldots$ enumerates the coordination spheres,
and the coefficients $\lambda_i$ fall rapidly with
increasing distance. 

\begin{center}
\begin{tabular}{|c|c|c|}\hline
 & $(i,j,k)$ & $\lambda$\\ \hline
$\lambda_0$ & (0,0,0)  & 0.935 \\
$\lambda_1$ & (1,0,0)  & -0.104 \\
$\lambda_2$ & (1,1,1)  & 0.056  \\
$\lambda_3$ & (1,1,0)  & -0.052 \\ \hline
\end{tabular}
\end{center}

TABLE 1. The coefficients $\lambda({\bf m} - {\bf n})$ where
${\bf m} - {\bf n} = i\hat{x}+j\hat{y}+k\hat{z}$.
\vspace{.2in}

By construction, the states $|0\rangle$ and $|\Psi_{ex}\rangle$ belong to
the set of eigenstates of the Hamiltonian $\tilde{H}$ (\ref{1.5})
with nonlocal terms
$\widetilde{H}_{1h} =
\sum_{{\bf m}\neq{\bf n},\sigma}t_{\bf mn}\tilde{c}^\dagger_{1,{\bf
m}\sigma}
\tilde{c}_{1,{\bf n}\sigma}$ and $\delta H_{12}$ (defined below) excluded.

However,  these nonlocal interactions
are responsible for forming the IV state. In a simplest approximation
we restrict ourselves by considering only the first term $\sim \lambda_0$
in the expansion (\ref{1.13}) and take the nonlocal interaction in the form
\begin{equation}
\delta H_{12}=\sum_{{\bf mn}\sigma}W({\bf m}-{\bf n})
f^\dagger_{{\bf m}\sigma}f_{{\bf n}\sigma}
f^\dagger_{{\bf n}\sigma}d_{{\bf m}\sigma} + {\mbox H.c.}
\label{1.14}
\end{equation}
This component of electrostatic interaction violates the point crystalline
symmetry and induces $fp$ hybridization at the site ${\bf m}$
in presence of a hole in the neighbouring cell ${\bf n}$.
The operator $\delta H_{12}$ has a property
$$\delta H_{12}|\psi_{\bf m}0_{\bf n}\rangle=
W({\bf m}-{\bf n})|0_{\bf m}0_{\bf n}\rangle
$$
where $0_{\bf n}$ stands for the cell ${\bf n}$ in the ground state
configuration. Thus this interaction intermixes the states
$|0\rangle\equiv\prod_{\bf n}|0_{\bf n}\rangle$
and $|\Psi_{ex}\rangle$. The mixing
constant is $w=\langle 0|\delta H_{12}|\Psi_{ex}\rangle=
zF(E_{ex})\lambda_0W$ where $z$ is the coordination number for Sm sublattice
and $W$ is the NN interaction matrix element.
After diagonalization the local neutral states in each cell ${\bf m}$
are represented by the linear combinations
\begin{eqnarray}
|\bar{0}_{\bf m}\rangle&=&\cos\theta |0_{\bf m}\rangle+
\sin\theta|\psi_{\bf m}\rangle, \nonumber \\
|\bar{\psi}_{\bf m}\rangle&=&-\cos\theta|\psi_{\bf m}\rangle+
\sin\theta|0_{\bf m}\rangle
\label{1.15}
\end{eqnarray}
where $\tan 2\theta\approx 2w/E_{ex}$,
and the valence is determined by the value of
$\sin^2\theta$.
As was mentioned, the local states (\ref{1.15})
are in some sense the electron-hole analogs
of two-hole Zhang-Rice singlets and triplets \cite{Zhang88,Belin94}.
Finally, the ground state of IV semiconductor is
\begin{equation}
|\Psi_0^{(iv)}\rangle=\prod_{\bf m}|\bar{0}_{\bf m}\rangle
\label{1.15a}
\end{equation}
and the low-lying local excitations are described by the vector
\begin{equation}
|\Psi_{ex}^{(iv)}\rangle=N^{-1}\sum_{\bf m}|\bar{\psi}_{\bf m}\rangle
\langle \bar{0}_{\bf m}|\Psi_0^{(iv)}\rangle.
\label{1.15b}
\end{equation}

The valence and the energy scale of these
local excitations (valence fluctuations) are characterised by the
degree of admixture of the excitonic state,
in which an electron is promoted to a
loosely bound ``molecular orbit" $P_{\bf m}$. If the exciton energy $E_{ex}$
is small enough (the exciton binding energy is comparable with the gap
width),
the value of $\sin^2\theta$ is close to 1/2.
Having in mind also the mean-field
part of hybridization given by Eq. (\ref{1.4}), we conclude that in this
case the valence 
\begin{equation}
n_v=3-\bar{n}_f+\sin^2\theta
\label{1.16}
\end{equation}
can exceed 2.5 both in SmS and SmB$_6$. Thus, the
large deviation of valence from the integer value and the softness
of local valence fluctuations are, apparently, the correlated phenomena.
Further discussion of internal consistency of the model can be found 
in the last section. 
\section{Trapping of Conduction Electrons and Hopping Conductivity}

The samples of good quality which were studied in the experiments of the
past
decade mentioned in the Introduction are $n$-type semiconductors at
low temperature, unlike
the $p$-type samples of the first generation
\cite{Nick71}. The electron concentration in these
samples is estimated at $n_e\sim 10^{17}$ cm$^{-3}$ at ambient pressure and
liquid
helium temperature \cite{Bat93,Cool95}. In order to interpret the transport
properties of these samples one should determine the spectrum of
electrons at the bottom of the conduction band in the presence of soft
valence fluctuations.
In our model the local valence fluctuations arise as transitions
between the states
$|\bar{0}_{\bf m}\rangle$ and $|\bar{\psi}_{\bf m}\rangle$
given by Eq. (\ref{1.15}).
It is known that these valence fluctuations
are the source of strong anomalies in the vibration spectra because
the characteristic time $\tau_{vf}$ of valence fluctuations is close to
phonon times $\tau_{ph} \sim \omega^{-1}_{ph}\sim 10^{-13}$s. Therefore,
one can expect that these ``slow" excitations could dress the carriers and
form
an electron-polaron cloud similar to the phonon cloud which results in
polaron self-trapping in dielectric crystals (see, e.g., \cite{Sumi73}).

To describe electron self-trapping
we start with a Hamiltonian including
the interaction between the conduction electron and the {\it local}
valence fluctuations in a single lattice site ${\bf n}=0$,
\begin{eqnarray}
H_e&=&\sum_{\bf k}\varepsilon({\bf k})c^{\dagger}_{\bf k}c_{\bf k}+
\Omega_0 A^\dagger A\nonumber \\
&&+ \sum_{{\bf k}_1,{\bf k}_2} \left[
W_{vf}({\bf k}_1,{\bf k}_2) c^{\dagger}_{{\bf k}_1}c_{{\bf
k}_2}A^{\dagger} + H.c.\right] .
\label{3.1}
\end{eqnarray}
This Hamiltonian stems from our basic Hamiltonian (\ref{1.5}). It
includes the electrons in the upper conduction band (the spin
summation and band index $s=2$ are omitted). Only the term which
corresponds to momentary, local redistributions of charge and
valency at the site of the excitation due to the interaction with the
charge carrier is retained in $H_{12}$. The valence fluctuations
with the energy $\Omega_0$ are described by the local operator
$A^\dagger=|\bar{\psi}\rangle\langle\bar{0}|$. This excitation is
in fact a charge transfer between the periphery and the centre of
the cell. The ``breathing" mode $\Omega_0$ describes the
polarisation of Sm ions in the cation sublattice that accompanies
the propagation of an excess conduction electron (predominantly,
over the samarium sublattice, see below).

The matrix element $W_{vf}({\bf k}_1,{\bf k}_2)$ is given by the integral
\begin{eqnarray*}
&&W_{vf}({\bf k}_1,{\bf k}_2)=  \\
 && \int d{\bf r}_1d{\bf r}_2
\bar{\psi}^*_{ex}({\bf r}_1)\bar{\psi}_{0}({\bf r}_1)
W({\bf r}_1,{\bf r}_2)
\psi^{*}_{{\bf k}_1}({\bf r}_2)\psi_{{\bf k}_2}({\bf r}_2)
\end{eqnarray*}
where $\psi_{{\bf k}}({\bf r})$ are the Bloch functions for the
conduction electrons,
$\bar{\psi}^*_{ex}({\bf r})$ and $\bar{\psi}_{0}({\bf r})$
are the wavefunctions of the states created by the operators
(\ref{1.15}). Their product is
\begin{eqnarray}
&&\bar{\psi}^*_{ex}({\bf r})\bar{\psi}_{0}({\bf r})
\equiv D({\bf r}) \nonumber \\
&&=
\frac{1}{2}\left(\sin 2\theta\left[
\rho_p({\bf r})-\rho_f({\bf r})\right]
-\cos 2\theta f_{xyz}({\bf r})d_{xyz}({\bf r})\right),
\label{3.1a}
\end{eqnarray}
where $\rho^f({\bf r})=|f_{xyz}(r)|^2$ is the f-electron density in
the centre of the cell (see Fig. 1) and
$\rho^p({\bf r})=F^2(E_{ex})\lambda_0^2|d_{xyz}(r)|^2$
is the $p$-electron density in a central cell given by the first term
in the r.h.s. of Eq. (\ref{1.13}).
The conduction electron density operator
$
\rho_{{\bf k}_1,{\bf k}_2}({\bf r}_1)=
\psi^{*}_{{\bf k}_1}({\bf r}_2)\psi_{{\bf k}_2}({\bf r}_2)
$
is determined by the Bloch functions
(\ref{1.4}). Near the bottom of conduction band this operator
can be approximately presented as 
\begin{equation}
\rho_{{\bf k}_1,{\bf
k}_2}({\bf r}_1)\approx \rho^f({\bf r})+\zeta_c\rho^p_{{\bf
k}_1,{\bf k}_2}({\bf r}_1) 
\label{3.2a}
\end{equation}
by means of Eq. (\ref{1.4d}). Here
\begin{equation}
\zeta_c=V_0^2/\varepsilon_{fb}^2,
\label{3.2b}
\end{equation}
\begin{equation}
 \rho^p_{{\bf
k}_1,{\bf k}_2}= \sum_{\langle{\bf j}\rangle_{NN}} 
e^{-i({\bf k}_1-{\bf k}_2)\cdot{\bf j}} |\psi_{{\bf j}}({\bf r}_1)|^2 
\label{3.2c}
\end{equation}
and $\varepsilon_{fb}={\varepsilon _{f}}-\varepsilon_{k_b}.$
Then the coupling constant in the Hamiltonian (\ref{3.1}) acquires the form
\begin{equation}
W_{vf}({\bf k}_1,{\bf k}_2)\approx w_0+
w_1\tilde{\beta}_{{\bf k}_1-{\bf k}_2}
\label{3.3}
\end{equation}
where
$$w_{0}=
\int d{\bf r}_1d{\bf r}_2D({\bf r}_1)W({\bf r}_1,{\bf r}_2)\rho^f({\bf
r}_2),$$

$$w_1\tilde{\beta}_{{\bf k}_1-{\bf k}_2}=
\int d{\bf r}_1d{\bf r}_2D({\bf r}_1)W({\bf r}_1,{\bf r}_2)
\rho^p_{{\bf k}_1,{\bf k}_2}({\bf r}_2+{\bf j}).
$$
The structure factor $\tilde{\beta}_{\bf k}$ is given by
\begin{equation}
\tilde{\beta}_{\bf k} =
8\cos\frac{k_x}{2}\cos\frac{k_y}{2}\cos\frac{k_z}{2}.
\label{beta2}
\end{equation}
It is seen that the first term $w_{0}$ dominates in the
electron-exciton interaction, therefore we begin with a calculation of
the electron self-energy induced by this term within the lowest order
of perturbation theory. The theory which allows inclusion of
nonlocal corrections due to small term $\sim w_1$ in Eq.
(\ref{3.3}) is described in Appendix. We calculate the electron
Green's function in Matsubara representation
\begin{equation}
G_{\bf{k k'}}(\omega_n)=\int_0^\beta\langle T_\tau
c_{\bf{k}}(\tau) c^{\dagger}_{\bf{k'}}\rangle e^{i\omega_n\tau}d\tau
\label{3.4}
\end{equation}
[$\omega_n=(2n+1)\pi T$] interacting with a localized valence
fluctuation mode $\Omega_0$ described by the propagator
\begin{equation}
D(\omega_m)=-\frac{2\Omega_0}{\omega_m^2+\Omega_0^2}
 \label{3.5}
\end{equation}
[$\omega_m=2\pi mT$]. The Dyson equation for the electron
Green's function is
\begin{equation}
G_{{\bf k},{\bf k}'}(\omega_n)  = G^0_{\bf
k}(\omega)\left[\delta_{{\bf k},{\bf k}'}  +
\sum_{{\bf k}_2}
\Sigma_{{\bf k},{\bf k}_2}(\omega_n)
G_{{\bf k}_2,{\bf k}'}(\omega_n)\right].
\label{3.6}
\end{equation}
In the lowest order of perturbation theory
the self-energy $\Sigma_{{\bf k}_1,{\bf k}'}(\omega_n) $
after carrying out an analytic continuation into the region
of real frequencies acquires the form (see Fig. 2)
\begin{equation}
\Sigma^R_{{\bf k},{\bf k}_2}(\omega)  = \sum_{{\bf k}_1}
W_{vf}({\bf k},{\bf k}_1)W_{vf}({\bf k}_1,{\bf k}_2) {\cal P}_{{\bf
k}_1}(\omega,\Omega_0) \label{3.7}
\end{equation}
where
\begin{eqnarray}
{\cal P}_{{\bf k}_1}(\omega,\Omega_0)&= &
\int_{-\infty}^{\infty}\frac{d\epsilon}{2\pi} \frac{{\rm
Im}D^R(\epsilon)} {\epsilon+\varepsilon({\bf k}_1)-\omega-i\delta}
\nonumber \\
&& \times \left( \tanh\frac{\varepsilon({\bf k}_1)}{2T}+
\coth\frac{\epsilon}{2T} \right) \label{3.8}.
\end{eqnarray}

\begin{figure}[ht]
\epsfysize=1.5in
\epsfbox[-60 -15 475 200]{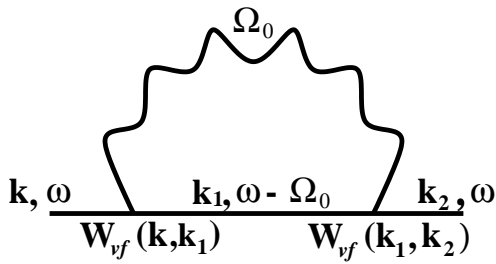}
\caption{The self-energy diagram corresponding to Eq. (37).  The straight
line represents a conduction electron and the wavy line represents a
soft valence fluctuation.
\label{fi:selfe}}
\end{figure}

Approximating the nonlocal potential $W_{vf}({\bf k},{\bf k}_1)$
by its local part $w_0$, we come immediately to the following
secular equation
\begin{equation}
1-w_0^2m(\omega)m(\omega-\Omega_0)=0 \label{3.10}
\end{equation}
where $m(\omega) = \sum_{\bf k}G^0_{\bf k}(\omega)$.

In close analogy with the polaron self-trapping effect
\cite{Sumi73}, one can expect that the attractive polarisation
potential with effective coupling constant
$$ g^0(\omega)=w_0^2\sum_{{\bf k}_2}
   G^0_{{\bf k}_2}(\omega-\Omega_0)$$
results in appearance of bound electron-polaron states. As
usual in 3D problems, the threshold value of the attractive
potential determines the onset of the bound state.
To find the constraints on the solutions with negative energy
(corresponding to bound states) we examine the function
$m(\omega)$  defined in (\ref{3.10}),
\begin{equation}
m(\omega) = \int_{-\pi}^{\pi}\frac{dk_x dk_y dk_z}{(2\pi)^3}
    \frac{1}{\omega-\epsilon({\bf k})}.
\label{3.10c}
\end{equation}
At $\omega=0$ the integral may be evaluated numerically (see Appendix).
Fig. 3
demonstrates the graphical solution of Eq. (\ref{3.10}) for a
reasonable set of model parameters. The value of $\Omega_0=5.5$meV
correlates the experimentally observed peak in optical
reflectivity spectra \cite{Ohta91,Nanba93}, and the binding energy
of localized $\omega_1\approx 3.5$meV  is in good agreement with
the activation energy registered in multiple optical and transport
measurements in the temperature interval 6 -- 14 K (see
\cite{Sluchan99} and references therein). 

We have demonstrated the existence of a self-trapped state in the
simplest approximation, whereby the local polarisation mode is
taken into account in  second order perturbation theory. However,
the values of the parameters necessary to achieve reasonable
agreement with experiment give for the dimensionless coupling
constant $\alpha=w_0/\Omega_0$ the value of $\alpha\approx 4.5$.
This means that in fact we are in a strong coupling limit, and a
more refined treatment is necessary. The generalisation of the
theory to the strong coupling limit can be done in close analogy
with the theory of small polaron. One can make the canonical
transformation 
$\tilde{c}_{\bf k}=e^{-S} c_{\bf k}e^{S}$, where
\begin{equation}
 S=\sum_{\bf kq}\frac{W_{vf}({\bf
k,q})}{\Omega_0} (c^{\dagger}_{\bf k}c_{{\bf k}+{\bf q}}A_{\bf
q}^{\dagger}- c^{\dagger}_{{\bf k}+{\bf q}}c_{\bf k}A_{-{\bf q}})
\label{3.10a}
\end{equation}
(see, e.g., \cite{Lang63}). We have seen above that the local term
$w_0$ is dominant in electron-exciton coupling constant
(\ref{3.3}). Then neglecting the contribution of the ``tail" $w_1$
we come to purely local interaction at a given site ${\bf m}={\bf 0}$,
namely, $\delta H_{12}= w_0 c^{\dagger}_{\bf 0}c_{\bf 0}(A_{\bf
0}^\dagger +A_{\bf 0})$. Eliminating this interaction by means of a
canonical transformation (\ref{3.10a}), one comes to the effective
Hamiltonian
\begin{eqnarray}
\tilde{H_e}  &=&  e^{-S}H_e e^{S} 
\nonumber \\
&=& \sum_{\bf m}
\varepsilon_f c^{\dagger}_{\bf m}c_{\bf m}  + \sum_{\bf m,n \neq
0}T_{\bf mn}c^{\dagger}_{\bf m}c_{\bf n}+ \Omega_0 A_{\bf
0}^\dagger A_{\bf 0}  \nonumber\\
&-&\varepsilon_{pol}c^{\dagger}_{\bf 0}c_{\bf 0}  -
\sum_{\bf m \neq 0}\left( T_{\bf m0}c^{\dagger}_{\bf 0}c_{\bf m}
e^{-\alpha
\left(A_{\bf 0}^\dagger -A_{\bf 0} \right)}+ \mbox{h.c.}\right).
\label{3.10b}
\end{eqnarray}

Now the main part of electron-exciton interaction is taken into
account exactly, and it is contained in the polaron shift. This shift
induces the local scattering at the site ${\bf 0}$ which is given by
the operator $\varepsilon_{pol}\sum_{{\bf k}_1,{\bf k}_2}c^\dagger_{{\bf
k}_1}c_{{\bf k}_2}$, where $\varepsilon_{pol}=\alpha w_0$.
The local scattering can be inserted in the self-energy of
the conduction electron in the same manner as it was done in second
order of perturbation theory in Eq. (\ref{3.6}). However, now the
calculation is exact, and instead of Eq. (\ref{3.10}) one has
\begin{equation}
0=1+\varepsilon_{pol}\sum_{\bf k}[\omega-\varepsilon({\bf k})]^{-1}
=1+\varepsilon_{pol}m(\omega)
\label{3.11}
\end{equation}
which can be solved by using the same approximation as in Eq.
(\ref{3.10}).  In any case the electron can be trapped by local
valence fluctuations provided the polaron shift
$\varepsilon_{pol}= \alpha w_0$ given by the bound solution of
Eq. (\ref{3.10}) exceeds the characteristic kinetic energy of the
electrons near the bottom of the conduction band. Therefore the
effective mass enhancement due to $pf$-hybridization favours the
formation of a self-trapped state.

\begin{figure}[ht]
\epsfysize=2.0in \epsfbox[40 350 575 650]{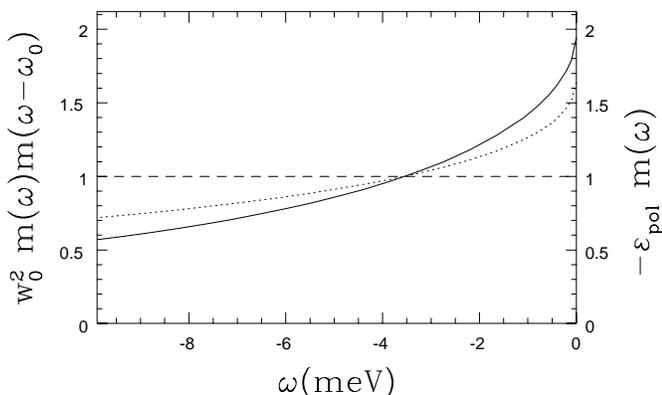}
\caption{Graphical solutions to Eqs. (39) (solid line) and (43)
(dotted line) using
$\epsilon_0 =
5$meV, $m^* = 100 m_0$, $\Omega_0=5.5$meV, $w_0 = 25$ meV and 
$\varepsilon_{pol} = 21$meV.
\label{fi:soln2}}
\end{figure}

In Fig. \ref{fi:soln2} we show a bound state
solution with energy 3.5 meV in agreement with the activation
energy observed in transport measurements and
absorption edges in optical experiments.   In both cases the coupling
constant $w_0$ is treated as a free parameter.  In the second order
perturbation calculation $w_0$ is determined to
be 25 meV, which as we have discussed above, indicates that our
solution lies in the limit of strong coupling.
In the non-perturbative calculation 
 we find that $\varepsilon_{pol} = \frac{w_0^2}{\Omega_0} = 21$ meV, which
implies that $w_0 = 11$ meV.  Because we are in the strong coupling
limit we do expect a deviation  between the
exact calculation and the
second order perturbation result.  However, our results are close
enough to suggest that the physical content of the perturbative calculation
is correct,  that is,  that an electron may bind to a single
local valence fluctuation and the result is a localised electron-valence
fluctuation complex.  Moreover,  we also observe that our results are
not particularly sensitive to the input parameters, and are therefore
not the result of any special ``fine tuning".
All of this is a consequence of the fact that the relevant energies
are all roughly the same:
$\Omega_0 = 5.5$ meV, $\hbar^2/a^2 m^* = 5$ meV and $\Delta_{act} = 3.5$
meV.

The last term in the effective Hamiltonian (\ref{3.10b}) is
responsible for the exciton-assisted hopping in an
electron-polaron band. The electron self-trapping can occur around any
arbitrary lattice site, so the trapped particle can move from a 
given cell to the neighboring cells by the mechanism which
resembles the polaron propagation \cite{Sumi73,Lang63}. Therefore,
the electron-polaron drift in electric field is responsible for
the conductivity at $T\to 0$ in n-type SmB$_6$. We leave
systematic treatment of electron transport in IV semiconductors
for future publications, and conclude this article by qualitative
discussion of anomalous low-$T$ behavior of SmB$_6$ mentioned
in the Introduction.

\section{Concluding remarks}

As a result of above analysis, one can understand 
the mechanism of reduction from the eV energy scale characteristic
of initial Hamiltonian (\ref{1.0}) down to meV scale of low-lying charge 
excitations in the $n$-type material. First, at the mean-field level of 
approximation (eqs. \ref{1.1} to \ref{1.2b}) the gap  $\Delta_0 \sim 10-20$ meV 
opens, and the states near the bottom of conduction band become heavy. 
This initial reduction of the energy scale is characterized by the parameter 
$\zeta_c$ (\ref{3.2b}), and the position of the $f$-level relative to the 
bottom/top of the bare conduction band $\epsilon_k$ can be extracted from 
the ratio of effective masses
\begin{equation}
r\equiv 
\frac{\varepsilon_{fb}}{\varepsilon_{ft}}
=\left(\frac{\zeta_v}{\zeta_c}\right)^{1/2}\approx
\left(\frac{m^*_c}{m^*_v}\right)^{1/2},
\label{C.01}
\end{equation}
where $\varepsilon_{ft}=\varepsilon_{k_t}-\varepsilon_{f}$,   
$\epsilon_{k_t}$ is the top of the band $\epsilon_k$, and the parameter 
$\zeta_v=(V_0/\varepsilon_{ft})^2$ is determined similarly to $\zeta_c.$ 
We neglect in these crude 
estimates the anisotropy of hybridization integral $V({\bf k})$. The hole 
effective mass $m^*_v$ is evaluated as $\sim 500-1000 m_0$ 
\cite{Wacht94,Sluchan99}, so we take $r \sim 1/3$, which is consistent with 
our assumption that the mean-field hybridization gives only part of the 
actual value of the valence $n_v$. 
According to our previous calculations \cite{Kik80}, the value of $r=1/3$ 
corresponds to $\bar{n}_f\approx 0.75$, and the experimentally observed value 
of $n_v\approx 2.55$ can be reached {\it only} by means of excitonic mechanism
with the exciton mixing parameter $\sin\theta \sim 0.55$ (see eq. \ref{1.16}). 
This value, in turn, 
agrees with our assumption of a soft excitonic mode.
Then, we find from (\ref{1.2b}) $\zeta_{c}=0.01$, and 
taking for the bandwidth 
$\varepsilon_{fb}+\varepsilon_{ft}$ the value of 4 eV in accordance with 
the band calculations for the related system \cite{Har88}, 
we estimate the hybridization coupling constant as $V_0\approx 0.1$ eV which 
is a reasonable value for the rare-earth materials. To make these estimates 
self-consistent, the hybridization gap should be found. The gap is defined 
via the above parameters as 
$\Delta_0\approx \zeta_c\varepsilon_{fb}+\zeta_v\varepsilon_{ft}$. Inserting the
values of corresponding parameters, we find $\Delta_0\approx 13$ meV which is 
in a reasonable agreement with the experimental data.   

Now, turning to the n-doped materials, we deal with heavy electrons, which  
interact with the valence fluctuations. The kinetic energy of these electrons 
is estimated as $\sim V_0^2/\varepsilon_{fb}\sim 10^{-2}$ eV , and the 
energy scale of valence fluctuations is, by their origin, limited from 
above by the energy gap  of the same $10^{-2}$ eV width.    
The polarization coupling constant $W_{vf}$ (\ref{3.3}) as well as the 
``superhybridization" matrix elements $W$ in the Hamiltonian (\ref{1.14}) should be
at least an order of magnitude less than the mean-field hybridization $V$, 
so the value of $w_0=25$ meV, which is used in our numerical solution of Fig. 3 
looks realistic. Eventually, solving the equation (\ref{3.11}) for the 
polaron shift, we descend one more step along the energy scale and find ourselves
in a meV region. The propagation of self-trapped electron-polaron can in 
principle be characterized by even lesser energies of the order of
$10^{-1}$ meV. 

Thus, we have found that the heavy electrons near the bottom of conduction band
can propagate in a lattice only in a polarization cloud of valence fluctuations.
It is worth mentioning that this conclusion correlates partly with a recent proposition 
of Kasuya \cite{Kas94} in spite of the fact that his model of the ground state 
of a mixed valence semiconductor disagrees with our picture in several respects.
He chooses the trial wave function for a local singlet (which he calls Kondo
singlet) of the form
\begin{equation}
|\tilde{0}_{\bf m}\rangle= \alpha_{f{\bf m}}|0_{\bf m}\rangle+
\alpha_{d{\bf m}}A^\dagger_{d{\bf m}}|\psi_{\bf m}\rangle
\label{C.02}
\end{equation}
where $A^\dagger_{d{\bf m}}=d^\dagger_{\bf m}c_{d{\bf m}}$ is a charge transfer
operator, which creates an electron in a 5$d$ shell of Sm ion and a hole of 
d-symmetry spread over surrounding boron sites similarly to our f-like orbital
$P_{\bf m}$ entering the state $|\psi_{\bf m}\rangle$ (\ref{1.13}). The physical 
reason of such a choice is conviction that the 5$d$ electrons also constitute a 
strongly correlated subsystem  and have a trend to form a bonding Kondo-like
state \`{a} la $f$ states in conventional Kondo lattices. However, looking at 
the real band structure of the rare-earth hexaborides, we see that the 
admixture of $d$ states to the band $\epsilon_k$ is really small: the center 
of gravity of $d$ states in divalent hexaborides is at the energy of 
$\sim 7-8$ eV above the bottom $\epsilon_{k_b}$ of conduction band (see, e.g., 
Fig. 5 in Ref. \cite{Mas97}). Therefore, 5$d$ levels are practically unoccupied,
and one cannot expect any kind of effective screening in a $d$ channel. 
Nevertheless, our model includes effective screening of {\it excess} electron 
in conduction band which resembles Kasuya's mechanism: the canonically 
transformed operators $\tilde{c}_{\bf 0}$ in a weak coupling limit can be 
presented as $\tilde{c}_{\bf 0}\sim c_{\bf 0}(1+\alpha A^\dagger_{\bf 0}),$
and the operator $A^\dagger_{\bf 0}$ returns part of the charge density back to the
Sm site from the periphery of the unit cell, i.e. plays the same role as the 
operator $A^\dagger_{d{\bf 0}}$ in the trial function (\ref{C.02}). 
The screening (or polaron dressing) in our model is due to the same 
valence fluctuations as all other physical effects, so the totality of 
experimental data is explained in a self-consistent 
scheme without appealing to
any additional hypotheses. 

As a result, our model is free from undesirable features of ``Kondo insulator"
approach which are not confirmed by the experiment. 
\noindent\\
(i) We do not appeal to the Kondo mechanism of forming the ground state singlet,
and the independence of the activation gap on external magnetic field
\cite{Aron95}, which rules out the Kondo insulator mechanism, agrees well with the 
expectations of our model.
\noindent\\
(ii) The ground state of our Hamiltonian is absolutely homogeneous, and this
statement agrees fairly with available experimental observations, 
whereas the trial 
function (\ref{C.02}) implies charge modulation in a form of 
Wigner crystal or Wigner liquid \cite{Kas94}. \noindent\\
(iii) The model of ref. \cite{Kas94} gives a single gap which predetermines 
electronic, optical and magnetic properties of the material, and it is unclear 
whether it is compatible with a real experimental situation which definitely 
evidences several energy scales for electronic and optical characteristics of
SmB$_6$ \cite{Sluchan99,Gorsh99}. \noindent\\
(iv) Unlike the bare ground state $|0\rangle$, 
the wavefunction (\ref{C.02}) is not fully symmetrical, so one can expect a
sort of ferroelectric ordering as $T\to 0$ \cite{Sham96,Kim95}.

Turning to the problem of low-temperature transport, we conclude  
that as $T\to 0$ the electron
propagation is valence fluctuation assisted motion of
in extremely narrow ``polaron band" separated by the gap
$\Delta_{act}$ from conduction band continuum. The residual
resistivity $\rho_0$ is inversely proportional to the hopping rate
between two neighboring crystal cells,
\begin{equation}
\rho_0^{-1}\sim S_{\bf
mn}\sim T_{\bf mn}\Phi_{\bf mn},
\label{c.1}
\end{equation}
 where
$\Phi_{\bf mn}=\langle \bar{\psi}_{\bf m}|\bar{\psi}_{\bf n}\rangle$ is a
function describing the overlap of the valence fluctuation
``clouds" centered around the sites  ${\bf m}$ and ${\bf n}$. As
$T\to 0$ the process is elastic, so the conductivity is
temperature independent. In SmB$_6$ this type of conductivity is
observed for $T<3 $K \cite{Bat93,Sluchan99,Gorsh99}. At higher
$T$ the hopping is, apparently, assisted by phonon and exciton
emission/absorption. Of course, the temperature
dependence of this hopping can differ from that for usual variable
range hopping in impurity bands. This transient regime can be seen
in the range 3 K $< T < $6 K, although the experimental data on the
temperature
dependence are still ambiguous \cite{Bat93,Sluchan99,Gorsh99}.
In the temperature interval 6 K $ < T < $14 K thermally
activated resistivity in SmB$_6$ with an activation energy of
$\Delta_{act} \approx 3.5 $meV is observed
\cite{Wacht94,Bat93,Cool95,Sluchan99,Gorsh99}.
At higher temperatures the electrons dissociate from their local valence
fluctuation clouds and as a result of this detrapping find
themselves at the bottom of the conduction band. At these temperatures
the activation of valence electrons also gives a significant contribution
to the electron conductivity.

The same three regimes (III, II and I, respectively, in terms of Ref.
\cite{Sluchan99}) with additional pronounced maximum around 5 K
are observed in the temperature behavior of the Hall
constant $R_H(T)$ \cite{Cool95}.
This maximum can be explained, at least qualitatively,
within a simple phenomenological picture of two groups of carriers
with high and low electron mobilities, $\mu_b=cR_b\sigma_b$ and
$\mu_h=cR_h\sigma_h$, where $R_{b,h}$ and $\sigma_{b,h}$ are
the Hall constants and conductivities of light $(b)$ and heavy (h)
carriers, respectively.
In the case of
hopping in impurity bands of doped semiconductors \cite{Efr84},
two contributions, $\sigma_h$
(hopping) and $\sigma_b$ (band) in the electron conductivity result in
the following equation for the Hall constant,
\begin{equation}
R_H=
\frac{R_b\sigma_b^2+R_h\sigma_h^2}{(\sigma_b+\sigma_h)^2}.
\label{c.2}
\end{equation}
Then
the maximum in $R_H(T)$ corresponds to a crossover from 
hopping motion at low $T$ to band motion at high $T$, provided
$\mu_b\gg \mu_h $.

In our case the phenomenological background of this equation still exists,
but the microscopic origin of all temperature dependences should be revised
because of the essentially many-particle nature of heavy carriers.

First of all, it is clear that the standard estimates of the number of
``scattering centers" obtained from the value of the residual resistivity
$\rho_0$ are simply inapplicable in our case, since we deal with
hopping of many-particle electron-exciton complexes
rather than with the motion of
extended band electrons. So, there is no room for the unitarity
limit arguments in these estimates, and the ``superunitarity
scattering" reflects the many-particle nature of current carriers
in semiconductors with fluctuating valence \cite{Varma91}. Thus,
the paradox of the number of scatterers per site \cite{Cool95}
is removed, and we return to the usual situation with one scatterer
per unit cell.

The next question is the enormously strong pressure dependence of
the residual resistivity  and the Hall coefficient \cite{Cool95}.
Again, according to our model,
one cannot directly apply the notions of charge transport in a
band of extended states for an estimation of the carrier concentration.
We think that the key to the extraordinary sensitivity of the
residual resistivity $\rho_0$
and $R_H(T\to 0)$ to external pressure is
{\em the increase of Sm valence with growing pressure} $P$. The eventual
source of this increase is growing $fp$-hybridization. Because of increasing
valence $n_v$ as a function of pressure
the matrix element $W_{vf}$ (\ref{3.3}) decreases.
As a result the polaron shift 
$\varepsilon_{pol}$ as well as the binding energy of
the self-trapped electron decreases too. The excitonic overlap function
$\Phi_{\bf mn}$ also grows due to lattice contraction. Since both the
hopping integral $T_{\bf mn}$ and the function $\Phi_{\bf mn}$ depend
exponentially on the intersite distance, one can expect a very sharp
dependence of the hopping rate on external pressure. The radius of the
localized state with binding energy 3.5 meV and effective mass
$\sim 100m_0$ is estimated to be  2-4
$\AA$ \cite{Sluchan99},
and an increase of this radius
by an order of magnitude would be enough to get a 10$^4$ growth of the
hopping
rate.  Such an increase is achievable under a pressure of 50 kbar. At higher
pressures
the trap becomes too shallow (the valence too close to 3)
to catch the electron,
the system transforms into conventional n-doped degenerate semiconductor,
and eventually it becomes a metal with trivalent Sm ions
in the cation sublattice.

Finally, the question of the variation of electron concentration $n_e$
with increasing $P$ and $T$ also demands special consideration.
Usually information about $n_e$ is extracted from the value of $R_H$ under
the
assumption of single band conductivity when
$R_H\approx R_b=(n_eec)^{-1}$.
According to the experimental data cited above, $R_H$ is nearly constant in
the temperature interval I (below 3 K), but falls drastically with
applied
pressure: $R_H$(45 kbar)/$R_H$(1 bar)$\sim 10^{-4}$.
However, in our case of two-component systems with
$\sigma_h\gg \sigma_b$ the Hall constant takes the form
$$
R_H \approx \frac{\mu_b\sigma_b+\mu_h\sigma_h}{c\sigma_h^2}.
$$
It is seen immediately that the pressure dependence of this function is
determined mainly by the denominator, and the decrease of $R_H$ with
pressure
is due to the exponentially growing factor $S_{\bf mn}$ in Eq. (\ref{c.1})
for $\sigma_h$ rather than to increasing $n_e$.

Generally speaking, $R_H$ derived from Eq. (\ref{c.2}) also cannot be used
to determine electron concentration neither in region I nor in
in region II.  The low $T$ limit of $n_{e0} \sim 10^{17}$cm$^{-3}$ 
which is obtained under the assumption that $R_H \approx R_b$  can
only be considered as a lower bound.  In the general case of two mechanisms
of charge transport the Hall constant is 
$R_H = (n_e e c)^{-1} \mu_H/\mu_D$,  where
$\mu_{H,D}$ are the Hall and drift  mobilities respectively.
If $\mu_H/\mu_D \gg 1$ at low $T$ the real electron concentration $n_e$
may substantially  exceed the value of $n_{e0}$.  Referring to the 
exponential dependence 
of $\sigma(T)$ and $R_H(T)$ in region II we should note that there are 
at least two causes of such a dependence. First there is the thermal 
activation  
of electrons from the polaronic traps to the band continuum states,
and second is the temperature dependence of the hopping rate
$S_{\bf mn}$ due to the contribution of thermally activated exciton
assisted terms which appear instead of the overlap integral  $\Phi_{\bf
mn}$.
As a result, both $\sigma_h(T)$ and $\sigma_b(T)$ give an
exponential contribution with the activation energy of $\Delta_{act}\sim
3.5$
meV
to the electron conductivity and the Hall constant (\ref{c.1}).
Again, it is impossible to calculate the variation of the electron
concentration $n_e(T)$ directly
from this equation. We leave the detailed evaluation
of various transport coefficients for a forthcoming paper,
but to conclude this
qualitative discussion we would like to emphasize that according to the
theory proposed in this paper the total carrier concentration in conduction
band is $n_e\sim  10^{17}$ cm$^{-3}$   at least while the
interband electron activation is negligible $(T\ll \Delta_0)$,
and there is no room for
dramatic pressure or temperature variation of $n_e$ in the mixed-valence
phase of $n$-SmB$_6$. The real semiconductor to metal transition occurs only
at
$P> P_c$ when the Sm valence changes to the integer value of +3.

The authors are grateful to M. Aronson, D.I. Khomskii and  N.E. Sluchanko
for useful discussions. This work was supported by Israeli Academy of
Sciences and 
Humanities Centre for ``Strongly Correlated Interacting Electrons in
Restricted Geometries".
S.C. acknowledges the support of the
Feinberg School of the Weizmann Institute.

\section*{Appendix}

We describe below the self-trapped state of the electron captured by
the nonlocal potential (\ref{3.3}). For the sake of simplicity we
consider the case of conduction band 
with a minimum at the $\Gamma$
point of the Brillouin zone.
To solve the Dyson equation (\ref{3.6})
we offer the procedure of factorization
of the vertex matrix element $W_{vf}({\bf k}_1,{\bf k}_2)$. The
latter can be represented in a form $W_{vf}({\bf k}_1,{\bf k}_2)
\equiv \sum_{a=0}^{8} \gamma^a_{{\bf k}_1}\gamma^a_{{\bf k}_2}$,
where
\begin{equation}
{\bf \gamma_ k} = \left(  \begin{array}{c}
                 \sqrt{w_0}   \\
     \sqrt{8\zeta w_1}\cos\frac{k_x}{2}\cos\frac{k_y}{2}\cos\frac{k_z}{2}
\\
     \sqrt{8\zeta w_1}\sin\frac{k_x}{2}\cos\frac{k_y}{2}\cos\frac{k_z}{2}
\\
     \sqrt{8\zeta w_1}\cos\frac{k_x}{2}\sin\frac{k_y}{2}\cos\frac{k_z}{2}
\\
     \sqrt{8\zeta w_1}\cos\frac{k_x}{2}\cos\frac{k_y}{2}\sin\frac{k_z}{2}
\\
    \sqrt{8\zeta w_1}\sin\frac{k_x}{2}\sin\frac{k_y}{2}\cos\frac{k_z}{2}  \\
     \sqrt{8\zeta w_1}\cos\frac{k_x}{2}\sin\frac{k_y}{2}\sin\frac{k_z}{2}
\\
     \sqrt{8\zeta w_1}\sin\frac{k_x}{2}\cos\frac{k_y}{2}\sin\frac{k_z}{2}
\\
     \sqrt{8\zeta w_1}\sin\frac{k_x}{2}\sin\frac{k_y}{2}\sin\frac{k_z}{2}
                 \end{array}  \right)
\label{tab1}
\end{equation}

By writing $G^a_{\bf k} \equiv \sum_{\bf k'}\gamma^a_{\bf k'}
   G_{{\bf k},{\bf k}^{'}}$
we obtain a solution for $G$ of the form
\begin{equation}
  G^a_{\bf k}(\omega) = (M^{-1})^{ab} \gamma^b_{\bf k}G^0_{\bf k}(\omega)
\label{gf}
\end{equation}
where
\begin{equation}
M^{ab}  = \delta^{ab} - \sum_{{\bf k}_1{\bf k}_2} \sum_{c=0}^8
G^0_{{\bf k}_1}(\omega)
   {\cal P}_{{\bf k}_2}(\omega,\Omega_0)
\gamma_{{\bf k}_1}^a\gamma_{{\bf k}_1}^c\gamma_{{\bf k}_2}^c
        \gamma_{{\bf k}_2}^b
\label{3.9}
\end{equation}
Here the index $a=0$ corresponds to the fully symmetric solution described
approximately by Eq. (\ref{3.10}), and $a=1$ stands for the nonlocal
contribution of the ``tail" $w_1$  with the same $A_1$ symmetry.
For the indices $a,b = 2\ldots 8$ the matrix $M$ is actually
diagonal since terms in the integrand must be even in
$k_{1x},k_{1y},k_{1z}$ and $k_{2x},k_{2y},k_{2z}$ and therefore
$a=c=b$. For these components $M$ is given by
\begin{equation}
M^{ab} =  \delta_{ab}\left[ 1-
m^a(\omega)m^a(\omega-\Omega_0)\right]
 \hspace{.3in} a,b= 2,\ldots 8
    \label{100}
\end{equation}
For the components $a,b=0,1$ $M$ reduces to a two by two matrix
\begin{equation}
M = \left(\begin{array}{cc}
         1-m^{1}(\omega)m^0(\omega-\Omega_0)
      & -m^0(\omega)\tilde{m}(\omega-\Omega_0) \\
  -  \tilde{m}(\omega)\tilde{m}(\omega-\Omega_0) &
   -\tilde{m}(\omega)m^1(\omega-\Omega_0) \\
&  \\
    - \tilde{m}(\omega)m^0(\omega-\Omega_0)
      & 1 - \tilde{m}(\omega)\tilde{m}(\omega-\Omega_0) \\
  - m^1(\omega) \tilde{m}(\omega-\Omega_0)  &
    - m^1(\omega) m^1(\omega-\Omega_0)
        \end{array}  \right) .
\label{101}
\end{equation}
In Eqs. (\ref{100}) and (\ref{101}) $m(\omega)$ is given by
\begin{eqnarray}
m^a(\omega) &=& \sum_{\bf k}
    G^0_{\bf k}(\omega)(\gamma_{\bf k}^a)^2 \hspace{.3in}
      a=0\ldots 8 \label{102a}\\
\tilde{m}(\omega) & = & \sum_{\bf k} G^0_{\bf
k}(\omega)\gamma_{\bf k}^0
       \gamma_{\bf k}^1.
\end{eqnarray}
Thus we
find that $M$ is reducible to the different representations  of
the group of symmetry operations on an octahedron $O_h$: $a=0,1$
and $a=8$ stand for the one-dimensional representations $A_1$ and
$A_2$ respectively, while $a=2,3,4,$ and $a=5,6,7,$ are the
indices of triplet states $T_1$ and $T_2$.

Hence, the secular equation which generalizes Eq. (\ref{3.10}) has the form
(for $\zeta \ll 1$)
\begin{eqnarray}
0 &= &\mbox{Det}M  \nonumber \\
  & = & [1-m^0(\omega)m^0(\omega-\Omega_0) - 2m^1(\omega)
    m^1(\omega-\Omega_0)] \nonumber  \\
  & & \times \prod_{a=2}^8\left[1-m^a(\omega)m^a(\omega-\Omega_0)\right]
\label{mw}
\end{eqnarray}
\begin{eqnarray}
m^{A_1,T_1,T_2,A_2}(\omega) = && 
 w_{1}\int_{-\pi}^{\pi} 
\frac{dk_x
dk_y dk_z}{(2\pi)^3}\nonumber \\
&& \hspace{-.3in}\frac{\zeta(1 \pm\cos k_x)(1 \pm \cos k_y)(1
\pm \cos k_z)}
     {\omega-\epsilon({\bf k})}~.
     \label{3.10aa}
\end{eqnarray}
Here the choice of three $+$signs corresponds to the $A_1$ state, two
$+$signs corresponds to the $T_1$ triplet, two $-$signs corresponds to
the $T_2$ triplet and three $-$signs corresponds to the singlet $A_2$.
To study
the analytical properties for small $\omega<0$, we use the fact
that the dominant contribution to the integral comes from small
values of ${\bf k}$. Near the bottom of conduction band
$\varepsilon_2({\bf k})$  the  hybridization is strong, and the
band is nearly flat [see Eq. (\ref{1.4})], so expanding the
dispersion around the minimum at the bottom, $\varepsilon(k)\approx
\hbar^2 k^2/2m^*$ in accordance with Eq. (\ref{1.2a}),
we refer to the heavy effective mass of
$m^*\approx 100 m_0$, observed experimentally \cite{Sluchan99,Gorsh99}. Then
we find for small $\omega < 0$,
\begin{eqnarray}
m^{0}(\omega) & \approx & \frac{w_0}{\epsilon_0}\left(-0.39 +
\frac{1}{2\pi}
         \sqrt{\frac{2|\omega|}{\epsilon_0}}\right) \\
m^{1}(\omega) & \approx & \frac{\zeta
w_{1}}{\epsilon_0}\left(-1.31 + \frac{1}{2\pi}
         \sqrt{\frac{2|\omega|}{\epsilon_0}}\right) \\
m^{T_1}(\omega) & \approx &  \frac{\zeta
w_{1}}{\epsilon_0}\left(-0.35 +
          0.28\frac{|\omega|}{\epsilon_0}\right)\\
m^{T_2}(\omega) & \approx &  \frac{\zeta
w_{1}}{\epsilon_0}\left(-0.20 +
         0.05\frac{|\omega|}{\epsilon_0}\right) \\
m^{A_2}(\omega) & \approx & \frac{\zeta
w_{1}}{\epsilon_0}\left(-0.14 + 0.02
      \frac{|\omega|}{\epsilon_0}\right)
\end{eqnarray}
where $\epsilon_0 = \hbar^2/a^2m^{*} \approx 5$meV.
Solutions  of Eq. (\ref{mw}) for the values of the parameters
given in Fig. 3 yield no bound states for $T_1$, $T_2$ and
$A_2$, so we conclude that the approximation $W_{vf}\approx w_0$
for the potential (\ref{3.3}) is sufficient for the description
of the bound electron-exciton states.

\end{document}